\documentclass[11pt,a4,tightenlines,nofootinbib]{revtex4}
\usepackage{amsmath,amscd,amssymb,epsfig}

\newcommand{\Vek}[1]{\mbox{\boldmath$#1$\unboldmath}}
\newcommand{\zr}[1]{\mbox{\hspace*{#1em}}}
\newcommand{\ID}{\mbox{{\sf 1}\zr{-0.16}\rule{0.04em}{1.55ex}\zr{0.1}}}

\begin{document}

\title{	Cosmic Strings Stabilized by Quantum Fluctuations}

\author{H. Weigel}

\affiliation{Physics Department, Stellenbosch University,
Matieland 7602, South Africa}

\begin{abstract}
We compute fermion quantum corrections to the energy of cosmic strings. A 
number of rather technical tools is needed to formulate this correction and 
we employ isospin and gauge invariance to verify consistency of these tools. 
These corrections must also be included when computing the energy of strings 
that are charged by populating fermion bound states in its background. We find 
that charged strings are dynamically stabilized in theories similar to the standard
model of particle physics.
\end{abstract}

\maketitle

\section{Introduction}
Various field theories suggest the existence of string--like configurations, 
which are the particle physics analogs of vortices or magnetic flux tubes. 
These configurations can arise at scales ranging from the fundamental distances 
in string theory to astrophysical distances, where in the latter case they are 
often called {\it cosmic  strings}, {\it cf.} 
Refs.~\cite{Copeland:2011dx,Hindmarsh:2011qj,Vilenkin:1994} for reviews.
A particularly interesting case is that of $Z$--strings, typically involving
the $Z$--boson field in theories similar to the standard model~\cite{Achucarro:1999it}. 
Since these string configurations are not classically stable we explore the
possibility that they are stabilized by quantum effects. The regularized 
and renormalized sum over the changes of all zero point energies of the 
quantum fluctuations in the string background, the so--called vacuum 
polarization energy (VPE), is central to these investigations. In field 
theory quantum effects are typically estimated by Feynman diagram
techniques. Unfortunately, string configurations have a non--trivial
structure at  spatial infinity which makes the formulation of a
Feynman perturbation expansion impossible without any further
adaptation. Even then, the convergence of the series is not guaranteed
as the relevant couplings are not  necessarily small and the series is
only asymptotic. Not surprisingly, the study of the VPE of cosmic 
string configurations has a long history of slow progress as reviewed 
in Ref.~\cite{Weigel:2015lva}. 

Starting point for parameterizing a cosmic string configuration is the $O(4)$
unit vector~\cite{Klinkhamer:1994uy,Graham:2006qt}
\begin{equation}
\hat{\Vek{n}}(\xi_1,\xi_2,\varphi)=\begin{pmatrix}
{\rm sin}\xi_1\,{\rm sin}\xi_2\, {\rm cos}\varphi,&
{\rm cos}\xi_1,& {\rm sin}\xi_1 \,{\rm cos}\xi_2,&
{\rm sin}\xi_1 \,{\rm sin}\xi_2\, {\rm sin}\varphi
\end{pmatrix}\,,
\label{eq:nhat}
\end{equation}
where the constant angles $\xi_1$ and $\xi_2$ are designated to describe the 
(weak) isospin orientation of the string
and $\varphi$ is the azimuthal angle in coordinate space.\footnote{The string 
configuration will be infinitely extended along the $z$--direction in 
coordinate space.} For simplicity, we will always consider unit winding of 
the string; generalizations to winding number $n$ merely require the replacement
${\rm cos}\varphi\to{\rm cos}(n\varphi)$ and ${\rm sin}\varphi\to{\rm sin}(n\varphi)$.

The unit vector $\hat{\Vek{n}} = (n_0,\Vek{n})$ defines the $SU(2)$ matrix
$U(\xi_1,\xi_2,\varphi)=n_0\ID-i\Vek{n}\cdot\Vek{\tau}$, where
$\Vek{\tau}=(\tau^1, \tau^2,\tau^3)$ are the three Pauli matrices.
The Higgs and gauge fields of the string are then characterized 
by two profile functions $f_H$ and $f_G$ that depend on the
distance $\rho$ from the center of the string:
\begin{align}
\begin{pmatrix} \phi_+(\rho,\varphi) \\[1mm] 
\phi_0(\rho,\varphi) \end{pmatrix}=
 f_H(\rho) U(\xi_1,\xi_2,\varphi)
\begin{pmatrix}0 \\[1mm] v \end{pmatrix}
\qquad {\rm and}\qquad
\Vek{W}(\rho,\varphi)= 
\frac{\hat{\Vek{\varphi}}}{g\rho} f_G(\rho)\,
U(\xi_1,\xi_2,\varphi)\,\partial_\varphi
U^\dagger(\xi_1,\xi_2,\varphi)\,.
\label{eq:profiles}\end{align}
Here, $v$ is the vacuum expectation value of the Higgs field that emerges from
spontaneous symmetry breaking and $g$ is the gauge coupling constant.
The gauge field $\Vek{W}$ is a vector in coordinate space and 
a matrix in the adjoint representation of weak iso--space. The profile functions 
vanish at the core of the string ($\rho=0$) and approach unity at spatial infinity. 

In what follows we will employ the abbreviations $s_i={\rm sin}\xi_i$ and 
$c_i={\rm cos}\xi_i$.  A global rotation within the plane of the second
and third component of $\hat{\Vek{n}}$ by the angle
$\alpha$ with ${\rm tan}\alpha=s_1c_2/c_1$ transforms it  into
\begin{equation}
\widetilde{\Vek{n}}(\xi_1,\xi_2,\varphi)=\begin{pmatrix}
s_1s_2 {\rm cos}\varphi,& \sqrt{1-s_1^2s_2^2},&
0,&  s_1s_2 {\rm sin}\varphi
\end{pmatrix}\,.
\label{eq:nhat1}
\end{equation}
Hence observables (which, by definition, are gauge invariant) will not
depend on the two angles $\xi_1$ and $\xi_2$ individually but only on the
product $s_1s_2$. Stated otherwise, all observables must remain invariant
along paths of constant $s_1s_2$ in isospin space~\cite{Klinkhamer:1997hw}.

\section{Fermion Vacuum Polarization Energy}

We focus on the fermion contribution to the VPE 
because for many (internal) fermion degrees of freedom, as {\it e.g.}
color, it dominates the boson counterpart. To be specific we consider an 
interaction between the fermions and the string background that is motivated 
by the standard model of particle physics. Introducing the matrix field
\begin{equation}
\Phi=\begin{pmatrix}
\phi_0^* & \phi_+ \cr -\phi_+^* & \phi_0 \end{pmatrix} 
\label{higgsmatrix}
\end{equation}
the model Lagrangian can be compactly written as
\begin{equation}
\mathcal{L}_\Psi=i\overline{\Psi}
\left(P_L {D \hskip -0.6em /}  + P_R {\partial \hskip -0.6em /} \right) \Psi
-f\,\overline{\Psi}\left(\Phi P_R+\Phi^\dagger P_L\right)\Psi\,.
\label{gaugelag}
\end{equation}
Here, $P_{R,L}=\frac{1}{2}\left(1\pm\gamma_5\right)$ are projection
operators on left/right--handed components, respectively and 
$D_\mu=\partial_\mu-ig\Vek{\tau}\cdot\Vek{W}_{\hspace{-0.1cm}\mu}$.
The strength of the Higgs-fermion interaction is parameterized by the 
Yukawa coupling $f$, which gives rise to the fermion mass, $m = f v$.

The configuration, Eq.~(\ref{eq:profiles}) approaches a local gauge 
transformation of the homogeneous vacuum configuration at spatial infinity. 
This creates the immediate problem that individual Fourier transformations 
(needed, {\it e.g.} to compute Feynman diagrams) of the Higgs and gauge fields 
are ill--defined. To avoid this problem we introduce an additional radial 
function $\xi(\rho)$ with the boundary values $\xi(0)=0$ and 
$\lim_{\rho\to\infty}\xi(\rho)=\xi_1$ to define the local $SU_L(2)$ gauge 
transformation
\begin{equation}
V={\rm exp}\left[-i\Vek{\tau}\cdot\Vek{\xi}(\rho,\varphi)\right]
\qquad {\rm with}\qquad
\Vek{\xi}(\rho,\varphi)=\xi(\rho)\,\begin{pmatrix}
s_2\, {\rm cos}\varphi \cr 
-s_2\, {\rm sin}\varphi \cr c_2
\end{pmatrix}\,.
\label{eq:localgauge}
\end{equation}
Since $\xi(0)=0$ this gauge transformation does not introduce any singularity 
at the origin from an un--defined azimuthal angle. At spatial infinity the
transformation accounts for the above mentioned gauge transformation of the 
constant vacuum.  We denote the Dirac Hamiltonian derived from the Lagrangian, 
Eq.~(\ref{gaugelag}) by $\mathcal{H}$ and transform it to 
$H=\left(P_R+VP_L\right)\mathcal{H}\left(P_R+VP_L\right)^\dagger$. For a 
compact presentation of $H$ we introduce $\Delta(\rho)\equiv\xi_1-\xi(\rho)$:
\begin{eqnarray}
H&=&-i\begin{pmatrix}0 & \Vek{\sigma}\cdot\hat{\Vek{\rho}} \cr
\Vek{\sigma}\cdot\hat{\Vek{\rho}} & 0\end{pmatrix} \partial_\rho
-\frac{i}{\rho}\begin{pmatrix}0 & \Vek{\sigma}\cdot\hat{\Vek{\varphi}}\cr
\Vek{\sigma}\cdot\hat{\Vek{\varphi}} & 0\end{pmatrix} \partial_\varphi
+\begin{pmatrix} 1 & 0 \cr 0 &-1\end{pmatrix}
+H_{\rm int}\,, \label{eqDirac0}\\[4mm]
H_{\rm int}&=&
\left[\left(f_H{\rm cos}(\Delta)-1\right)
\begin{pmatrix} 1 & 0 \cr 0 &-1\end{pmatrix}
+if_H\,{\rm sin}(\Delta)\begin{pmatrix}0 & 1 \cr -1 & 0\end{pmatrix}
I_H\right]
+\frac{1}{2}\frac{\partial \xi}{\partial \rho}
\begin{pmatrix}-\Vek{\sigma}\cdot\hat{\Vek{\rho}}
& \Vek{\sigma}\cdot\hat{\Vek{\rho}} \cr
\Vek{\sigma}\cdot\hat{\Vek{\rho}}
& -\Vek{\sigma}\cdot\hat{\Vek{\rho}}\end{pmatrix}I_H
\nonumber \\[3mm]
&&\hspace{1cm}+\frac{s_2}{2\rho}\, \begin{pmatrix}
-\Vek{\sigma}\cdot\hat{\Vek{\varphi}}
& \Vek{\sigma}\cdot\hat{\Vek{\varphi}} \cr
\Vek{\sigma}\cdot\hat{\Vek{\varphi}}
& -\Vek{\sigma}\cdot\hat{\Vek{\varphi}}\end{pmatrix}
\Big[f_G\,{\rm sin}(\Delta)\,I_G(\Delta)
+(f_G-1)\,{\rm sin}(\xi)\,I_G(-\xi)\Big]\,.
\label{eqDirac}
\end{eqnarray}
The isospin matrices in this expression are 
\begin{equation}
I_H=\begin{pmatrix}
c_2 & s_2 {\rm e}^{i\varphi} \cr 
s_2 {\rm e}^{-i\varphi} & -c_2 \end{pmatrix}
\qquad {\rm and} \qquad
I_G(x)=\begin{pmatrix}
-s_2{\rm sin}(x) & [c_2{\rm sin}(x)-i{\rm cos}(x)]\,{\rm e}^{i\varphi} \cr
[c_2{\rm sin}(x)+i{\rm cos}(x)]\,{\rm e}^{-i\varphi} &
s_2{\rm sin}(x)\end{pmatrix}\,.
\label{eq:IG}
\end{equation}
Note that $I_G$ appears with different arguments in Eq.~(\ref{eqDirac}).
Nothing from the invariance along the path with $s_1s_2={\rm const.}$ is
manifest in Eq.~(\ref{eqDirac}), nor is the gauge invariance from 
Eq.~(\ref{eq:localgauge}).

The eigenvalues of $H$ determine the VPE whose formal expression
\begin{equation}
E_{\rm vac}=\frac{m^2}{2\pi} \int_0^\infty d\tau\, \tau
\left\{\sum_{\ell} D_{\ell}\left[\nu(\tau,\ell)-\nu_1(\tau,\ell)-\nu_2(\tau,\ell)\right]
-\frac{c_F}{c_B}\sum_{\ell}\bar{D}_\ell\bar{\nu}_2(\tau,\ell)\right\}
+E_2+E_{\rm f.b.}
\label{eq:master}
\end{equation}
has been derived previously, {\it cf.} Ref.~\cite{Graham:2011fw}. Here 
$\nu(\tau,\ell)$ is the logarithm of the determinant of the fermion scattering
matrix (obtained from $H$) about the string analytically continued to imaginary 
momenta $ik=t=\sqrt{\tau^2-m^2}$.  Furthermore $\nu_{1,2}$ are the associated 
first and second order Born terms (obtained by iterating $H_{\rm int}$),
respectively. Their subtraction removes the dominating quadratic divergence.
The final subtraction under the integral arises from the 
second order Born term for scattering a complex boson field. The coefficients
$c_F$ and $c_B$ are computed such that the remaining logarithmic divergences
in the two orbital momentum sums with degeneracies $D_{\ell}$ and 
$\bar{D}_{\ell}$ cancel. Finally, $E_2+E_{\rm f.b.}$ are finite combinations
of Feynman diagrams that compensate the subtractions under the integral and 
counterterms that are unique for prescribed renormalization conditions. We 
stress that Eq.~(\ref{eq:master}) does not contain any (numerical) cut--off.

\section{Numerical Results}

We parameterize the string profiles as 
\begin{equation}
f_H(\rho)=1-{\rm e}^{-\frac{\rho}{w_H}}\,,\qquad
f_G(\rho)=1-{\rm e}^{-\left(\frac{\rho}{w_G}\right)^2}
\qquad{\rm and}\qquad
\xi(\rho)=\xi_1\left[1-
{\rm e}^{-\left(\frac{\rho}{w_\xi}\right)^2}\right]
\label{eq:para}
\end{equation}
and compute the VPE as a function of the width parameters 
${w_H}$, ${w_G}$ and ${w_\xi}$. We measure all length 
variables in multiples of $1/m$.

We first test our numerical calculations against the isospin invariance of 
Eq.~(\ref{eq:nhat1}) and the local gauge transformation of Eq.~(\ref{eq:localgauge}). 
The former implies  identical VPEs for all values of $\xi_1$ and $\xi_2$ with 
equal $s_1s_2$, while the latter requires constant VPEs as we vary $w_\xi$. 
Though the individual entries for the VPE in Eq.~(\ref{eq:master}) are not 
invariant by themselves, the final result must be. Since the local counterterms are 
manifestly invariant, it is sufficient to perform the calculation in the 
$\overline{\rm MS}$ renormalization scheme. The left panel of table \ref{tab1} 
shows that this VPE is the same for different values of the angles $\xi_1$ and $\xi_2$ 
that have equal $s_1s_2=0.29389$. To see that the obtained small variations are merely 
within numerical errors we also present the sums of the moduli that change 
dramatically. We see a similar independence of the gauge profile parameter $w_\xi$ 
in the right panel. Thus we have confirmed the required symmetries.
\begin{table}[b]
\parbox[r]{8.4cm}{\small \begin{tabular}{c|c|c|c||c||c}
$\xi_1/\pi$ &$\xi_2/\pi$ & $E_{\delta}$ &  $E_{\rm FD}$ & $E_{\rm vac}$
& $|E_{\delta}|+|E_{\rm FD}|$\cr
\hline
0.1 & 0.4 & 0.1504 &  0.0014 &~~0.1518~~~& 0.1518\cr
0.4 & 0.1 & 0.1702 & -0.0180 &~~0.1521~~~& 0.1882 \cr
0.3 & 0.11834 & 0.1496 & 0.0021 &~~0.1517~~~& 0.1517 \cr
0.2 & 1/6 & 0.1639 & -0.0117 &~~0.1522~~~& 0.1758
\end{tabular}}
\parbox[l]{7.4cm}{\small \begin{tabular}{c|c|c||c||c}
{$w_\xi$} & $E_{\delta}$ &  $E_{\rm FD}$ & $E_{\rm vac}$
& $|E_{\delta}|+|E_{\rm FD}|$\cr
\hline
2.0 & 0.3010 & -0.0108 &~~0.2902~~~& 0.3118 \cr
3.5 & 0.2974 & -0.0072 &~~0.2902~~~& 0.3046 \cr
5.0 & 0.2953 & -0.0047 &~~0.2905~~~& 0.3000 \cr
6.5 & 0.2915 & -0.0015 &~~0.2901~~~& 0.2930
\end{tabular}}
\caption{\label{tab1}Isospin (left panel, $w_H=w_G=w_\xi=3.5$) 
and gauge (right panel, $w_H=w_G=4.82$, $\xi_1=0.3\pi$, $\xi_2=0.25\pi$) 
invariances~\cite{Weigel:2016ncb}. Reference to Eq.~(\ref{eq:master}):
$E_\delta$ is the $\tau$--integral contribution and 
$E_{\rm FD}=E_2+E_{\rm f.b.}$.}
\end{table}

\begin{figure}
\centerline{
\epsfig{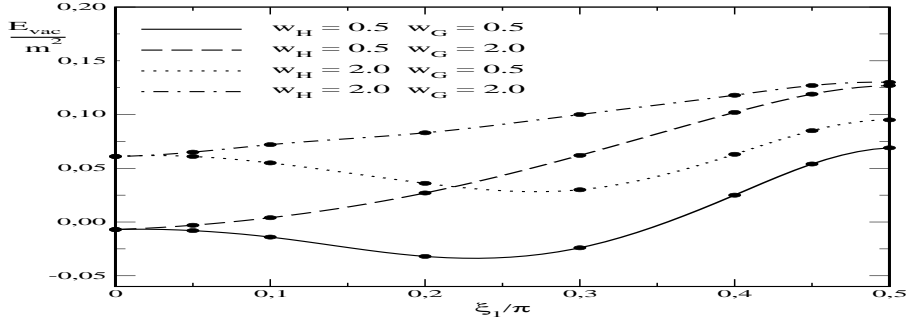}}
\caption{\label{fig1}VPE results in on--shell renormalization from 
Ref.~\cite{Graham:2011fw}.}
\end{figure}

In the next step we compute the VPE as a function of the variational parameters
$w_H$ and $w_G$ with on--shell renormalization conditions. In view of the above 
established invariances, we may choose particular values for $\xi_1$, $\xi_2$ 
and $w_\xi$. For numerical stability we take the latter similar to $w_H$ and 
$w_G$. Furthermore we take $\xi_2=\pi/2$ since it renders $H$ real thereby 
simplifying the scattering problem. Then the three variational parameters 
are $w_H$, $w_G$ and $\xi_1$. Typical results are shown in figure~\ref{fig1}.
We find the VPE to be positive for most values of the variational parameters,
so that it does not contribute to binding. For narrow string configurations 
(small width parameters) negative results are indeed obtained. For them 
to overcome the classical energy ($\mu_H=m_H/m$ is the scaled Higgs mass)
\begin{equation}
\frac{E_{\rm cl}}{m^2}=2\pi\int_0^\infty \rho d\rho\,\left\{
n^2s^2_1s^2_2\,\biggl[\frac{2}{g^2}
\left(\frac{f_G^\prime}{\rho}\right)^2
+\frac{f_H^2}{f^2\rho^2}\,\left(1-f_G\right)^2\biggr]
+\frac{f_H^{\prime2}}{f^2}
+\frac{\mu_h^2}{4f^2}\left(1-f_H^2\right)^2\right\}
\label{eq:classical}
\end{equation}
and bind the string, large Yukawa coupling 
constants are needed which eventually brings the Landau ghost problem into the 
game. Thus such bound configurations should not be considered~\cite{Graham:2011fw}.

\section{Charged Strings}

In particular wide strings (which are not affected by the Landau ghost problem)
generate many fermion bound states. Their energy eigenvalues are of the 
same order in the semi--classical $\hbar$ expansion as the VPE. Hence the 
inclusion of these levels ultimately enforces the consideration of the VPE.

Populating these levels can thus produce
a configuration with finite charge (per unit length) $Q$ whose total energy per
unit length is less than $Qm$. Subtracting the latter gives the 
total binding energy (also per unit length)
\begin{equation}
E_{\rm tot}(Q)=E_{\rm cl}+E_{\rm vac}+
\frac{1}{\pi}\sum_i\int_0^{p^F_i(Q)}dp \,
\left[\sqrt{\epsilon_i^2+p^2}-m\right]\,,
\label{eq:ebind}
\end{equation}
where $\epsilon_i$ and $p^F_i(Q)$ are the energy eigenvalues of $H$ with
$0\le\epsilon_i<m$ and the corresponding Fermi momenta, respectively. 
For a prescribed charge $Q$ we find an upper bound of $E_{\rm tot}(Q)$ 
by scanning several hundred configurations that are parameterized by 
different values of $w_H$, $w_G$ and $\xi_1$.

\begin{figure}
\centerline{
\epsfig{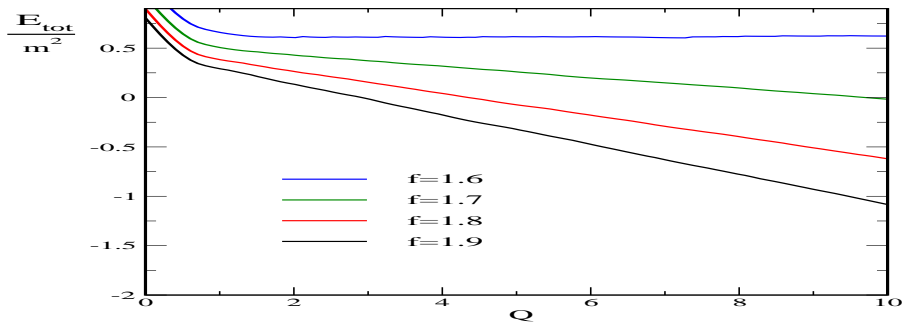}}
\caption{\label{fig2}(Color online) The total binding energy, Eq.~(\ref{eq:ebind})
for various values of the Yukawa coupling ($f=1.6,1.7,1.8,1.9$ top to 
bottom) from Ref.~\cite{Graham:2011fw}. The gauge coupling $g$ is taken at its
standard model value.}
\end{figure}

Numerical results for the upper bound of $E_{\rm tot}(Q)$ are shown in figure~\ref{fig2}.
For $f>f_c\approx1.7$ the total binding energy turns indeed negative and we have succeeded
in constructing a bound string configuration~\cite{Weigel:2010zk}. This critical 
value is only about twice as big as the Yukawa coupling of the top quark in the 
standard model.

\section{Conclusion}

We have developed a procedure to compute the VPE of cosmic strings with spectral
methods~\cite{Graham:2009zz}. We have verified this procedure numerically by reproducing 
the required invariances even though the individual entries of the calculation are not 
invariant by themselves. We have found that the VPE does not bind classically unstable 
string configurations. When charging the string by populating fermion bound states, a 
stable configuration is obtained for fermions only about twice as heavy as the top quark.
This must be considered a novel solution in a standard model like theory.

\acknowledgments
The author would like to thank the organizers of the workshop on {\it Strong Field 
Problems in Quantum Theory} for this worthwhile event. The author also very much
appreciates considerable contributions to this project from the collaborators 
N. Graham and M. Quandt. 

\noindent
The work is supported in part by the NRF (South Africa) by grant~77454.

\end{document}